\documentclass[aps,preprint,11pt]{revtex4}
\usepackage{amsmath,amssymb,amsthm,color}
\usepackage{epsfig}
\usepackage{graphicx}
\usepackage{amssymb}
\usepackage{amsfonts}
\usepackage{amsmath}
\usepackage{graphicx}%
\usepackage{subfigure}
\usepackage{setspace}

\newcommand{\ket}[1]{|#1\rangle}
\newcommand{\bra}[1]{\langle #1|}

\def\be#1{\begin{equation}\label{#1}}
\def\ee{\end{equation}}
\def\bea#1{\begin{eqnarray}\label{#1}}
\def\eea{\end{eqnarray}}

\def\tit#1{\textit{#1}}

\def\r{\right)}
\def\l{\left(}
\def\ket#1{|#1\rangle}
\def\bra#1{\langle#1|}
\begin{document}
\begin{spacing}{1.2}
\title{Reduction of entanglement degradation and teleportation improvement in Einstein-Gauss-Bonnet gravity}
\author{B. Nasr Esfahani }\email{ba_nasre@sci.ui.ac.ir}\affiliation{Department of Physics, Faculty of
Sciences, University of Isfahan , Isfahan 81744, Iran}
\author{M. Shamirzai}\email{mshj@iaush.ac.ir}\affiliation{Department of Physics, Faculty of
Sciences, University of Isfahan , Isfahan 81744, Iran}
\author{M. Soltani}\email{msoltani@phys.ui.ac.ir}\affiliation{Department of Physics, Faculty of
Sciences, University of Isfahan , Isfahan 81744, Iran}

\begin{abstract}

Bipartite entanglement for states of a non-interacting bosonic or
fermionic field in the spacetime of a spherically symmetric black
hole of Einstein-Gauss-Bonnet gravity, is investigated. Although
the initial state is chosen to be maximally entangled as the Bell
states, the Hawking-Unruh effect causes the state to be mixed and
the entanglement degrades, but with different asymptotic behaviors
for the fermionic and bosonic fields. The Gauss-Bonnet term with
positive $\alpha$ can play an anti-gravitation role and so this
causes to decrease the Hawking-Unruh effect and consequently
reduces the entanglement degradation. On the other hand, the
suggested higher dimensions for the spacetime, lead to more
entanglement degradation by increasing the dimension. There is a
dramatic difference between the behaviors of the entanglement in
terms of the radius of the horizon for a five-dimensional black
hole and that for higher dimensional black holes.
\end{abstract}

\maketitle


\section{Introduction}

More and more efforts have been expended on the study of quantum
entanglement, which is the essential resource of quantum
information processing, in relativistic setting for not only
logical completeness but also for the study of the physical bounds
of quantum information processing tasks. While Lorentz
transformations can  not change the overall quantum entanglement
of a bipartite state \cite{ging}, the situation for non inertial
observers is different. In order to investigate the
observer-dependent character of entanglement, the bipartite
entanglement for states of a non-interacting massless scalar field
when one of the observers is uniformly accelerated is studied by
Alsing and Milburn \cite{Alsing1}. They showed that the fidelity
of teleportation between two parts in relative acceleration
degrades by increasing  the acceleration. The bipartite
entanglement was also studied by Fuentes-Schuller and Mann, and
they showed that only inertial observers agree on the degree of
entanglement, and non-inertial observers see a degradation
\cite{Mann1}. The acceleration of the observer effectively
introduces  an environmental decoherence caused by the Unruh-Davis
effect. Also, Alsing \textit{et al} discussed the entanglement by
different modes of a fermionic field \cite{Mann2}. Their results
showed that different types of fields will have qualitative
different effects on the degradation of the entanglement. The
entanglement by different helicity modes of an electromagnetic
field in non inertial reference frame studied in \cite{Lin}.

Some authors extended this issue to the entanglement in curved
spacetimes, motivated by the fact that the spacetime near the
event horizon of a Schwarzschild black hole resembles Rindler
coordinates in the infinite acceleration limit \cite{Xian1}. Ahn
showed that for a two-mode squeezed state in a Riemannian
geometry, an initial Gaussian state becomes decoherent due to the
Hawking effect and in addition a  higher squeezing  leads to a
higher degradation \cite{Ahn1}.  In this way, the study of
entanglement in the black hole geometry is directly related to the
black hole information paradox \cite{Hawking}. Hawking radiation
in the background of an asymptotically flat static black hole in
Einstein gravity is investigated in Ref. \cite{Pan}. The same
issue for sonic black holes is covered in Ref. \cite{Ge2}. Quantum
no-cloning theorem for charged black holes is discussed in Ref.
\cite{Xian2}. Entanglement in a dynamical spacetime is
investigated for bosonic and fermionic fields  in Refs.
\cite{Ball} and \cite{Mann4}, respectively . The results show that
a dynamic metric can generate entanglement and conversely the
entanglement encodes information concerning the underlying
structure of spacetime. In principle it is possible to fully
reconstruct the parameters of the cosmic history from the
entanglement entropy. Also,  as is  shown in Ref. \cite{Mann5}, it
is possible to generate entanglement by accelerated observers
using the  Unruh mechanism. Ge extended the teleportation in
gravitational fields to the higher dimensional spacetimes such as
Schwarzschild and Kerr black hole solutions of the higher
dimensional Einstein theory of gravitation \cite{Ge-2008}.

The possibility that spacetime may have more than four dimensions
is now a standard assumption in high energy physics
\cite{Polchinski}. The idea of brane cosmology that is consistent
with string theory, suggests that matter and gauge interactions
may be localized on a brane embedded into a higher dimensional
spacetime such that the gravitational field can propagate in the
whole of the spacetime. In this way we need to consider gravity in
dimensions higher than four. In this context one may use another
consistent higher dimensional theory of gravity with a more
general action, that is the  Lovelock theory of gravitation which
contains higher powers of Riemann tensor and its derivatives
\cite{Lov}. The first and the second terms in the field equation
are the cosmological constant and the Einstein tensor,
respectively. The next term, which contains curvature-squared
terms is the Gauss-Bonnet tensor. Up to this order, the obtained
field equations are  called the Einstein-Gauss-Bonnet
gravitational field equations. Many Authors have obtained various
solutions for  these equations by  assuming some symmetries for
the metric. In Ref. \cite{Dehghani2} asymptotically AdS solutions
of Gauss-Bonnet gravity are obtained without cosmological
constant. Also it is shown that an accelerating universe can be
obtained from the modified Friedman equation in Gauss-Bonnet
gravity \cite{Dehghani3}. Therefore, it seems that the
Gauss-Bonnet term can have an anti-gravitation role.

As a further step in the subject of quantum information in higher
dimensional curved spacetimes, we will provide an analysis of
quantum entanglement for quantum fields in the spacetime of black
hole solution of Einstein-Gauss Bonnet theory. It can be
interesting to study how the Hawking temperature can change the
entanglement and teleportation in this spacetime. Remembering that
the Gauss-Bonnet gravity can have an anti-gravitation role, we
expect that in this improved gravitational theory,  the quantum
information tasks be enhanced. Moreover, we can investigate the
effect of higher dimensions on the behavior of the entanglement.
The out line of this paper is as follows. In Sec. \ref{secUnruh},
we briefly revisit the Gauss-Bonnet gravity and then review the
Hawking-Unruh effect in the black hole solution of this theory. In
Sec. \ref{secCAl}, we set up the problem and  calculate the
entanglement monotone for both fermionic  and bosonic  fields.
Fidelity of teleportation for a desired observer is derived and
explained  via appropriate figures in Sec. \ref{secFID}. Some
concluding remarks are given in Sec. \ref{fin}.

\section{Thermal distribution of Quantum States in Einstein-Gauss-Bonnet
black hole}\label{secUnruh}

As we know the field equations in Einstein-Gauss-Bonnet gravity
can be written as
\begin{equation}  \label{field}
      G^{(\textrm{E})}_{\mu \nu }+\Lambda g_{\mu \nu }+\alpha G_{\mu \nu
      }^{(\textrm{GB})}=T_{\mu \nu }
\end{equation}
where $T_{\mu \nu }$ is the energy-momentum tensor of matter,
$G^\textrm{E}_{\mu \nu }$ is the Einstein tensor   and   $G_{\mu
\nu }^{(\textrm{GB})}$ is the second order Lovelock tensor or
Gauss-Bonnet tensor, which is defined in terms of the curvature
tensor $R_{\mu\nu\sigma\kappa}$ as
\begin{eqnarray}
      G_{\mu \nu }^{(\textrm{GB})} &=&2(R_{\mu \sigma \kappa \tau }R_{\nu }^{\phantom{\nu}%
      \sigma \kappa \tau }-2R_{\mu \rho \nu \sigma }R^{\rho \sigma
      }-2R_{\mu \sigma }R_{\phantom{\sigma}\nu }^{\sigma }+RR_{\mu \nu
      })  \notag
      \label{Gaussfield} \\
      &&-\frac{1}{2}\left( R_{\mu \nu \sigma \kappa }R^{\mu \nu \sigma
      \kappa }-4R_{\mu \nu }R^{\mu \nu }+R^{2}\right) g_{\mu \nu
      },
\end{eqnarray}
and $\alpha$ is the Gauss-Bonnet constant which  we take it
positive. Let us consider a $d-$dimensional $(d\ge 5)$ static
spherically symmetric spacetime with the metric
\begin{equation}\label{metric}
     \mathrm{d}s^2 = -f(r)\mathrm{d}t^2+\frac{1}{f(r)}\mathrm{d}r^2+r^2
      \mathrm{d}\Omega_{d-2}^2,
\end{equation}
where $f(r)$ is an unknown function and
$r^2\mathrm{d}\Omega_{d-2}^2$ is the metric of a
$(d-2)$-dimensional subspace. It can be proved that this metric
describes a black hole solution of the field equations
(\ref{field}) with $\Lambda=0$, provided that
\begin{equation}\label{fd}
      f(r)=k+\frac{r^2}{2(d-3)(d-4)\alpha}\left(1\pm\sqrt{1+\frac{4(d-3)(d-4)\alpha
      m}{r^{(d-1)}}}\right)
\end{equation}
where $m$ is the geometrical mass of the black hole and $k$
denotes the curvature of the $(d-2)$-dimensional subspace
\cite{Dehghani3}. Of course, for the special case of $d=5$ the
function takes a particular form as
\begin{equation}\label{f}
     f(r)= k+\frac{r^{2}}{4\alpha }\pm \sqrt{\frac{r^{4}}{%
     16\alpha ^{2}}+\left( \left| k\right| +\frac{m}{2\alpha }\right)},
\end{equation}
which has a geometrical mass $m+2\alpha |k|$. Since we are
interested in asymptotically flat solutions, we must choose the
minus sign and also $k=1$ in (\ref{fd}) and (\ref{f}). It is easy
to show that in the limit of small $\alpha$, this $f(r)$ gives the
metric of a $d$-dimensional Schwarzschild solution of Einstein
theory, as a requirement of Lovelock gravitation theory.
Evidently, the radius of the horizon denoted by $r_h$ can be
obtained as the positive root of $f(r)$. For example in the case
of $d=5$ one simply obtains  $r_h=\sqrt{m}$.

\begin{figure}
     \begin{center}
        \includegraphics[width=10cm,height=8cm,angle=0]{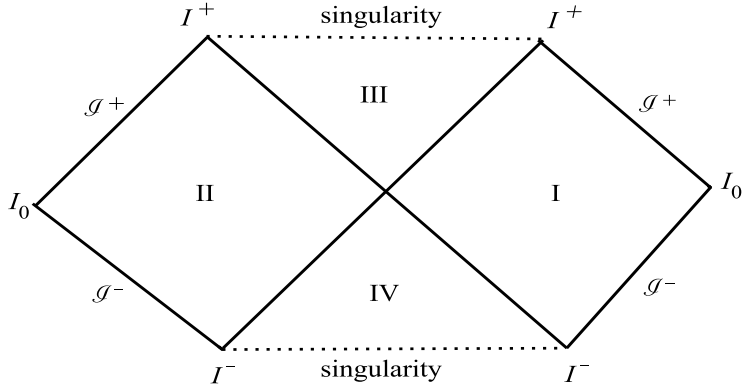}
      \caption{Penrose diagram for the solution (\ref{metric}).}\label{pen}
    \end{center}
\end{figure}

It is convenient to consider the Penrose diagram for the spacetime
(\ref{metric}). Therefore, we invoke the Kruskal coordinates by
using the appropriate coordinate transformations. For performing
this, we write the metric in $(u,v)$ coordinates as
\begin{eqnarray}\label{krs}
      &&ds^2 = - f(r)du dv+r^2d\Omega^2_{d-2},\nonumber\\
      &&u =t-r_{*}, \quad v=t+r_{*},
\end{eqnarray}
where the Regge-Wheeler tortoise coordinate is defined by
$r_*=\int{\mathrm{d}r/f(r)}$. We note that $r_*$ has a logarithmic
singularity at the event horizon. This behavior of tortoise
coordinate near the event horizon is generally establish for the
GR theory \cite{Nature}, and for the Schwarzschild spacetime in
this theory, we can explicitly expand $r_*$ around the radius of
the horizon. For the spacetime of Gauss-Bonnet black hole, we can
still expand $r_*$ in the neighborhood of $r_h$ as
 \be{rtor}
      r_*\approx
     \Gamma \ln(r-r_h)+\mathcal{G}(r-r_h)
 \ee
where $\mathcal{G}(r-r_h)$ is a nonsingular function at $r_h$.
Comparing the coefficient  $\Gamma$ with the analogous coefficient
in the Schwarzschild spacetime, we infer that this is exactly the
inverse of the Hawking temperature of Gauss-Bonnet black hole,
that is $\Gamma^{-1}=T$. This temperature can be defined
geometrically in terms of the surface gravity of black hole as
 \be{TH}
           T=2\pi\left(-\frac{1}{2}\nabla_{\mu}\xi_{\nu}\nabla^{\mu}\xi^{\nu}\right)^{-\frac{1}{2}},
 \ee
which as applied for the spacetime (\ref{metric}), leads to
 \be{THGB}
          T=\frac{1}{4\pi}\frac{d f(r)}{ dr}\Big|_{r_h}.
 \ee
This as evaluated for (\ref{fd}) and (\ref{f}), leads to
 \be{htd}
     T=\frac{1}{4\pi}\,{\frac
      {\alpha\,{d}^{3}-12\,\alpha\,{d}^{2}+47\,\alpha\,d+d\,{r_h}^{2
      }-60\,\alpha-3\,{r_h}^{2}}{r_h \, \left(
      2\,\alpha\,{d}^{2}-14\,\alpha \,d+24\,\alpha+{r_h}^{2} \right) }}
 \ee
for the $d$-dimensional spacetime, and
 \be{ht5}
     T=\frac{1}{2\pi}\,{\frac {r_h}{ \left( 4\,\alpha+r_h^2
     \right)}}.
 \ee
for the special case of $d=5$.

Now we define the Kruskal coordinates
\begin{equation}\label{UV}
          U\propto\pm e^{-uT}, \qquad V\propto\mp e^{vT}
\end{equation}
which are used for an analytical extension of the metric. The
upper (lower) sign  refers to the region I (II) in Fig.
(\ref{pen}) which represents the  Penrose diagram of the metric
(\ref{metric}). Notice that the regions I and II are causally
disconnected.

We consider two observers, who are  going to communicate through a
quantum information protocol in the spacetime of
Einstein-Gauss-Bonnet black hole. They share a Bell state, which
is composed  of two modes of a free  quantum field, say ground
state and the first exited state of a scalar or a spinor field.
These two observers meet  each other and  share the prescribed
quantum state at the asymptotic  region of the black hole. Then,
one of them, say Alice, stays on a timelike geodesic of the black
hole, and consequently is freely falling as an inertial observer.
But the other observer, Rob, approaches the event horizon, but he
barely accelerates to avoid of falling  in the black hole. In
order to express the entanglement between these observers, we
require to construct the quantum field modes as seen by each of
observers.

First we consider a  massless scalar field $\phi$ that satisfies
the Klein-Gordon equation
 \be{KG}
        \frac{1}{\sqrt{-g}}\partial_{\mu}(\sqrt{-g}g^{\mu\nu}\partial_{\nu}\phi)=0.
 \ee
where $g$ denotes the metric determinant. Regarding the spherical
symmetry of the metric, $\phi$ can be separated as
\begin{equation}
    \phi(t,r,\Omega) = e^{-i\omega t} \frac{R_{\omega
    l}(r)}{r^{\frac{d-2}{2}}} Y_{lm}(\Omega),
\end{equation}
where $Y_{lm}(\Omega)$ is the $(d-2)$-dimensional spherical
harmonic functions and $R_{\omega l}(r)$ satisfies the following
equation:
\begin{equation}\label{radial}
     \frac{\partial^2 R_{\omega l}}{\partial
    r_{*}^2}+\omega^{2}R_{\omega
    l}-f(r)\left(\frac{(d-2)^2}{4r^2}f(r)+\frac{d-2}{2r}\frac{df}{dr}+\frac{l(l+d)}{r^2}\right)R_{\omega
    l}=0.
\end{equation}

The observer Alice who is freely  falling into the black hole,
sees nothing special at the horizon.  Hence she has access to the
entire of the spacetime. But as Fig. \ref{pen} shows, for the
accelerated observer Rob, there are two causally disconnected
regions of spacetime denoted by I and II . Since the future
directed timelike killing vector corresponding to the region II is
directed in the opposite direction  of that of the region I, that
is $[\partial_t]_{I}=[-\partial_{t}]_{II}=[\partial_{-t}]_{II}$,
then the positive frequency solutions related to the regions I and
II differ up to minus sing in $t$. Indeed the positive frequency
solutions of Eq. (\ref{radial}) are obtained as
 \begin{eqnarray}\label{Analytic}
   \phi_{\textrm{I},k}&\sim &e^{ikr_*-i\omega t}\equiv e^{i\omega u}\\ \nonumber
   \phi_{\textrm{II},k}&\sim &e^{ikr_*+i\omega t}\equiv e^{-i\omega v}\\ \nonumber
 \end{eqnarray}
In this way any quantum field can be quantized as
 \be{p11}
     \Phi=\sum_{l,m}\int {d}\omega \left[\left(a_{\textrm{I}}\phi_{\textrm{I},k}+
     a_{\textrm{II}}\phi_{\textrm{II},k}\right)+H.C.
   \right]
 \ee
where $a_{\textrm{I},k}$ ($a_{\textrm{II},k}$) and
$a_{\textrm{I},k}^{\dagger}$ ($a_{\textrm{II},k}^{\dagger}$) are
the annihilation and creation operators for the mode $k$ in the
region I (II). These operators are called the Schwarzschild
operators and satisfy the following relations
\begin{eqnarray}
      a_{\textrm{I},k} \vert 0 \rangle_{\textrm{I},k} \otimes \vert n \rangle_{\textrm{II}} &=&
      a_{\textrm{II},k} \vert n \rangle_{\textrm{I}} \otimes \vert 0 \rangle_{\textrm{II},k} = 0,
       \nonumber \\  a^{\dagger }_{\textrm{I},k} \vert 0 \rangle_{\textrm{I},k} \otimes
       \vert n
       \rangle_{\textrm{II}} &=& \vert 1 \rangle_{\textrm{I},k} \otimes \vert n \rangle_{\textrm{II}},\nonumber \\
      a^{\dagger }_{\textrm{II},k} \vert n \rangle_{\textrm{I}} \otimes \vert 0
     \rangle_{\textrm{II}} &=& \vert n \rangle_{\textrm{I}} \otimes \vert 1
    \rangle_{\textrm{II},k}.
\end{eqnarray}

Since the solutions (\ref{Analytic}) can not be analytically
continued from  the region I to the region II, we must express
them in the Kruskal coordinates. Regarding Eqs. (\ref{UV})
and(\ref{Analytic}), we can show that the analytical solutions in
the whole of the spacetime can be written as \cite{BD}
 \bea{pp}
      \phi_{K,k}^{+}&=&e^{\frac{\pi\omega}{2T}}\phi_{\textrm{I},k}
      +e^{-\frac{\pi\omega}{2T}}\phi^{*}_{\textrm{II},-k},\nonumber \\
      \phi_{K,k}^-&=&e^{-\frac{\pi\omega}{2T}}\phi^*_{\textrm{I},-k}
      +e^{\frac{\pi\omega}{2T}}\phi^{}_{\textrm{II},k},
 \eea
which correspond to positive and negative frequencies with respect
to the Killing vector $\partial_U$. Instead of (\ref{p11}), one
can now expand $\Phi$ in terms of $\phi_{K,k}^+$ and
$\phi_{K,k}^-$ as
\begin{equation}\label{p1}
      \Phi=\sum_{l,m}\int {d}\omega \left[\left(b_{K,k}^+\phi_{K,k}^++b_{K,k}^-\phi_{K,k}^-\right)+H.C.
      \right]
\end{equation}
where
 \bea{aa}
        b_{K,k}^+&=&(\cosh{\eta}) a_{\textrm{I},k}-(\sinh{\eta}) a_{\textrm{II},-k}^{\dagger} \nonumber \\
       b^{-\dagger}_{K,-k} &=&(\sinh{\eta})a_{\textrm{I},k}^{\dagger}-(\cosh{\eta})a_{\textrm{II},-k}
 \eea
where $\tanh{\eta}=e^{-\pi\omega/T}$. These relations that can be
obtained by using Eqs. (\ref{pp}), are just the Bogoliubov
transformations between the Schwarzschild operators and the
Kruskal operators. The Kruskal vacuum and first excited states for
a known $k$ can be expressed as
\begin{eqnarray}\label{zero}
       \vert 0 \rangle_{K} = \frac{1}{\cosh \eta}\sum^{\infty}_{n=0}
       (\tanh^{n}\eta) \vert n \rangle_{\textrm{I}} \otimes \vert n \rangle_{\textrm{II}},
\end{eqnarray}
\begin{eqnarray}\label{one}
       \vert 1 \rangle_{K} &=& b_{K,k}^+ \vert 0
       \rangle_{K} \nonumber\\
       &=& \frac{1}{\cosh ^2 \eta}\sum^{\infty}_{n=0}( \tanh^{n}\eta )\sqrt{n+1}
        \vert n+1 \rangle_{\textrm{I}} \otimes \vert n\rangle_{\textrm{II}}.
\end{eqnarray}

For a Fermionic  quantum field,  we can repeat this argument to
obtain the appropriate Bogoliubov transformations as
 \bea{Bogfer}
         b_{K,k}^+&=&(\cos{\zeta}) a_{\textrm{I},k}-(\sin \zeta)a^{\dagger}_{\textrm{II},-k}\nonumber \\
          b_{K,-k}^{-\dagger}&=& (\sin{\zeta})a_{\textrm{I},k}+(\cos \zeta)a_{\textrm{II},-k}^{\dagger}
\eea
 where $\tan{\zeta}=e^{-\pi\omega/T}$. The ground state and
the first excited states of the field in Kruskal and Schwarzschild
coordinates are related by
\begin{eqnarray}
     \vert 0 \rangle_{K} &=& (\cos{\zeta})\vert 0 \rangle_{\textrm{II}} \otimes \vert 0
      \rangle_{\textrm{I}}+(\sin{\zeta})\vert 1 \rangle_{\textrm{II}} \otimes
       \vert1  \rangle_{\textrm{I}}\\
       \nonumber
      \vert 1 \rangle_{K} &=& \vert 1
     \rangle_{\textrm{I}}  \vert 0\rangle_{\textrm{II}}.
\end{eqnarray}

\section{Bipartite entanglements }\label{secCAl}

\subsection{Bosonic entanglement}

According to the previous section, we focus on the condition that
Rob  moves  toward the black hole and then stops  on a surface
outside the event horizon  by  a slow acceleration, and Alice is
freely  falling toward  the black hole and perhaps after a finite
proper time crosses  the event horizon. If they share a maximally
entangled Bell state far from the black hole as
\begin{equation}\label{bell}
       \vert \phi \rangle_{M} =
       \frac{1}{\sqrt{2}}\left(\vert 0 \rangle_{M}^{{~\textrm{A}}} \vert
        0 \rangle_{M}^{{~\textrm{R}}}
        +\vert 1 \rangle_{M}^{{~\textrm{A}}}\vert 1 \rangle_{M}^{{~\textrm{R}}}\right),
\end{equation}
where the first qubit in each term refers to Alice's cavity and
the second qubit refers to Rob's cavity  and the index $M$
indicates that the states are considered  in the Minkowski
spacetimee. The state inside the Rob's cavity is no longer
perfectly entangled with that of Alice due to the Unruh effect.
One can assume that prior to their coincidence, Alice and Rob have
not any particle in their cavities and each cavity supports two
orthogonal states with the same frequency (single mode
approximation), which each is excited to a single particle state
Fock state at the coincidence point.

If Rob undergoes a uniform acceleration or stays in a curved
spacetime, the state in his cavity must be specified in
Schwarzschild coordinates. As a consequence, the second ket  in
each term of (\ref{bell}) must be expanded according to
(\ref{zero}) and (\ref{one}).  We can then rewrite Eq.
(\ref{bell}) in terms of Minkowski modes for Alice and
Schwarzschild modes for Rob, which leads to a tripartite density
matrix as $\rho_{\textrm{A, I, II}}$. Since Rob is causally
disconnected from region II, we must trace over the states in this
region, then we obtain a mixed bipartite density matrix denoted as
$\rho_{\textrm{A,I}}$. By rearrangement of elements of the reduced
density matrix operator, this can be recast in a block diagonal
form as
\begin{equation}\label{leak}
       \rho_{\textrm{A,I}}=\frac{1}{2 \cosh^2\eta}\sum_{n}( \tanh^{2n} \eta)\rho_{n},
\end{equation}
where
\begin{equation}\label{rho}
       \rho_{n} = \vert 0,n \rangle \langle 0,n \vert
       +\frac{\sqrt{n+1}}{\cosh \eta} \vert 0,n \rangle \langle 1,n+1 \vert
       +\frac{\sqrt{n+1}}{\cosh \eta} \vert 1,n+1 \rangle
       \langle 0,n \vert + \frac{{n+1}}{\rm cosh^2 \eta}\vert 1,n+1 \rangle
       \langle 1,n+1 \vert,
 \end{equation}
and $\vert n,m \rangle = \vert n \rangle_{M}^{{~\textrm{A}}} \vert
m \rangle_{\textrm{I}}^{~\textrm{R}}$. Here the summation goes
over all values of $n$ as a consequence of the Bose-Einstein
statistics of the scalar field.

The degree of entanglement for the two observers  can be
quantified  by using the the concept of logarithmic negativity
\cite{vidal} defined as
\begin{equation}
          \mathcal{N}=\log_{2}{||\rho^T||}
\end{equation}
where $||\rho^T||$ is the trace norm of the partial transposed
matrix $\rho^T$, which is defined as the sum of the eigenvalues of
$\sqrt{(\rho^T)^\dag\rho^T}$. For a symmetric matrix it can be
shown that this is equal to the sum of the absolute value of the
eigenvalues of $\rho^T$. In this case the logarithmic negativity
vanishes unless some negative eigenvalues exist. If only one
negative eigenvalue $N$ exists, then the logarithmic negativity
can be rewritten as
\begin{equation}\label{LNform}
       \mathcal{N}=  \log_{2}(1-2N).
\end{equation}
The partial transpose of the density operator $\rho_{A,I}$ in Eq.
(\ref{leak}) can be obtained by interchanging the Alice's qubit,
such that the (n,n+1) block of this partial transposed matrix can
be written as
\begin{equation}\label{AIT}
      \frac{\tanh^{2n}\eta}{2 \cosh^2\eta}
       \left(
       \begin{array}{lc}
      \tanh^2{\eta} & \frac{\sqrt{n+1}}{\cosh \eta} \\
     \frac{\sqrt{n+1}}{\cosh \eta} & \tanh^{-2}{\eta}\frac{{n}}{\cosh^2{\eta}}
    \end{array}
\right).
\end{equation}
Then the negative eigenvalue of $\rho_{A,I}^{T}$ is obtained as
$N=\sum_n N_n$ where
\begin{equation}\label{nar}
      {N}_{n}=\frac{\tanh^{2n}\eta}{4
       \cosh^2\eta}\left(\frac{n}{\sinh^2{\eta}}+\tanh^2{\eta}-
        \sqrt{\left(\frac{n}{\sinh^2{\eta}}+\tanh^2{\eta}\right)^2+\frac{4}{\cosh^2{\eta}}}\right),
\end{equation}
is the negative eigenvalue of (\ref{AIT}). Now substituting $N$ in
Eq. (\ref{LNform}), we obtain the logarithmic negativity as an
infinite series. This logarithmic negativity is a function of
$\eta=\tanh^{-1} \left( {\rm e}^{-\pi \,\omega/T}\right)$ which
ranges from 0 to $\infty$. For a given $\omega$, the limit
$\eta\rightarrow 0$ corresponds to $T=0$ and as we can see from
Eq. (\ref{nar}), this leads to $\mathcal{N}=1$. On the other hand,
the limit $\eta\rightarrow \infty$, which corresponds to infinite
Hawking temperature, leads to $\mathcal{N}=0$ or complete
destruction of the entanglement.

\begin{figure}
\subfigure[\, Logarithmic negativity versus the Hawking
temperature. The entanglement vanishes asymptotically. ]{
    \label{BLNT}     \begin{minipage}[b]{0.3\textwidth}
      \centering
      \includegraphics[width=5cm,height=5cm]{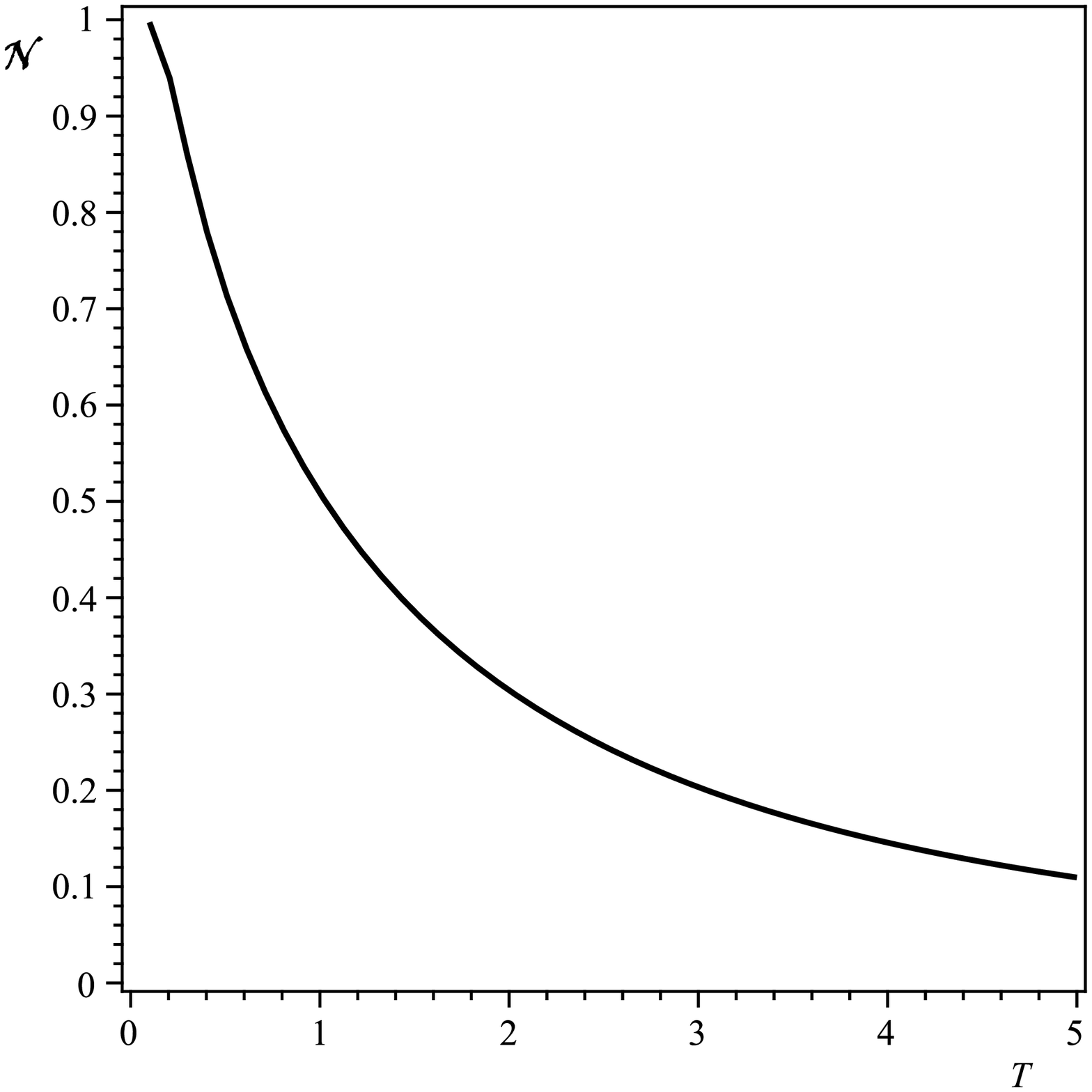}
    \end{minipage}}
\subfigure[\, Logarithmic negativity versus the spacetime
dimensions for some given $\alpha$. By increasing $\alpha$, the
entanglement becomes less sensitive to increase of $d$.]{
    \label{BLn}     \begin{minipage}[b]{0.3\textwidth}
      \centering
      \includegraphics[width=5cm,height=5cm]{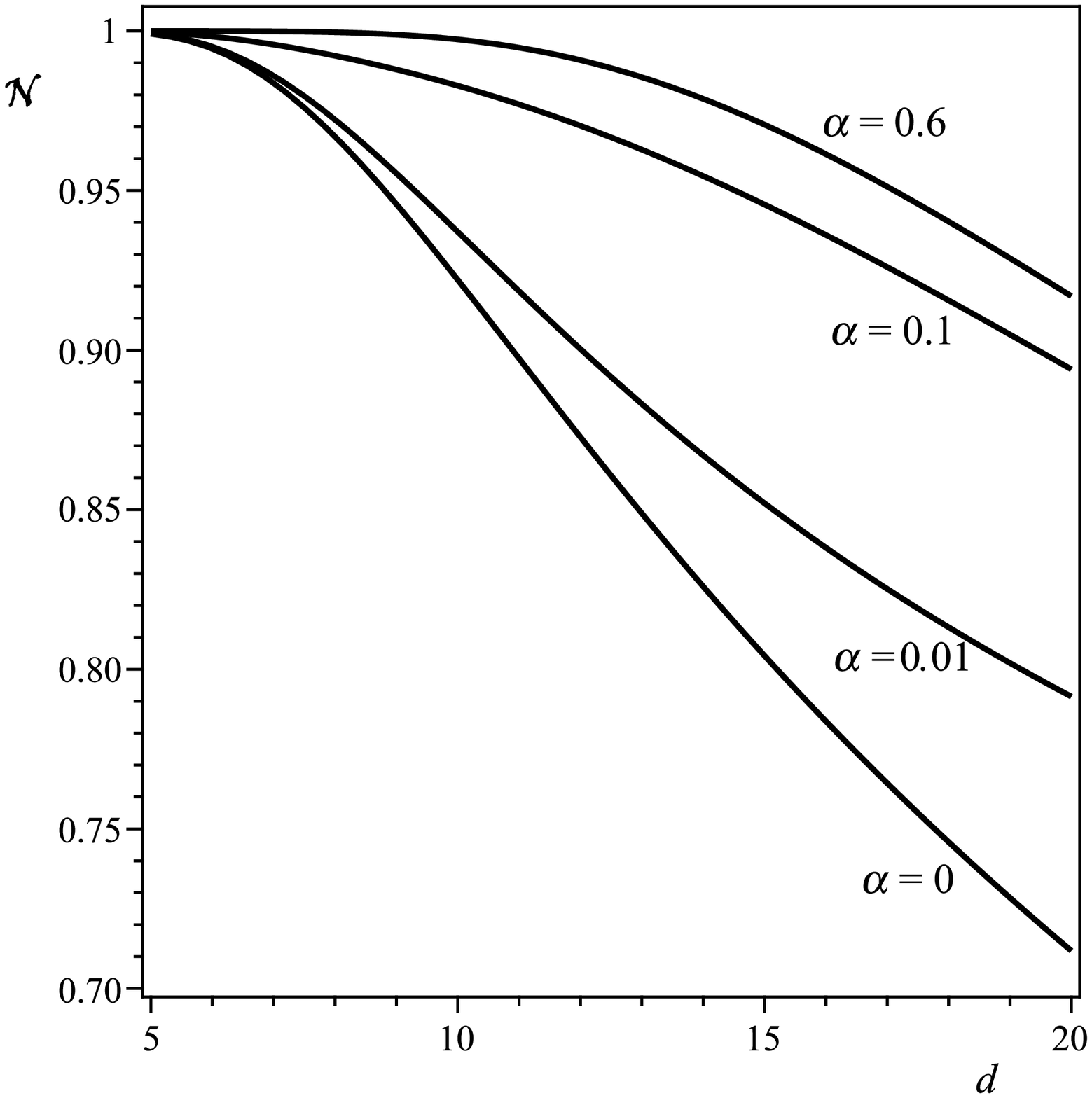}
    \end{minipage}}
\subfigure[\, Logarithmic negativity versus the Gauss-Bonnet
coefficient $\alpha$ for some given $d$. By increasing $\alpha$,
the entanglement grows; however, only for $d=5$ this can reach to
the maximal entanglement of 1.]{
    \label{BLAlpha}     \begin{minipage}[b]{0.3\textwidth}
      \centering
      \includegraphics[width=5cm,height=5cm]{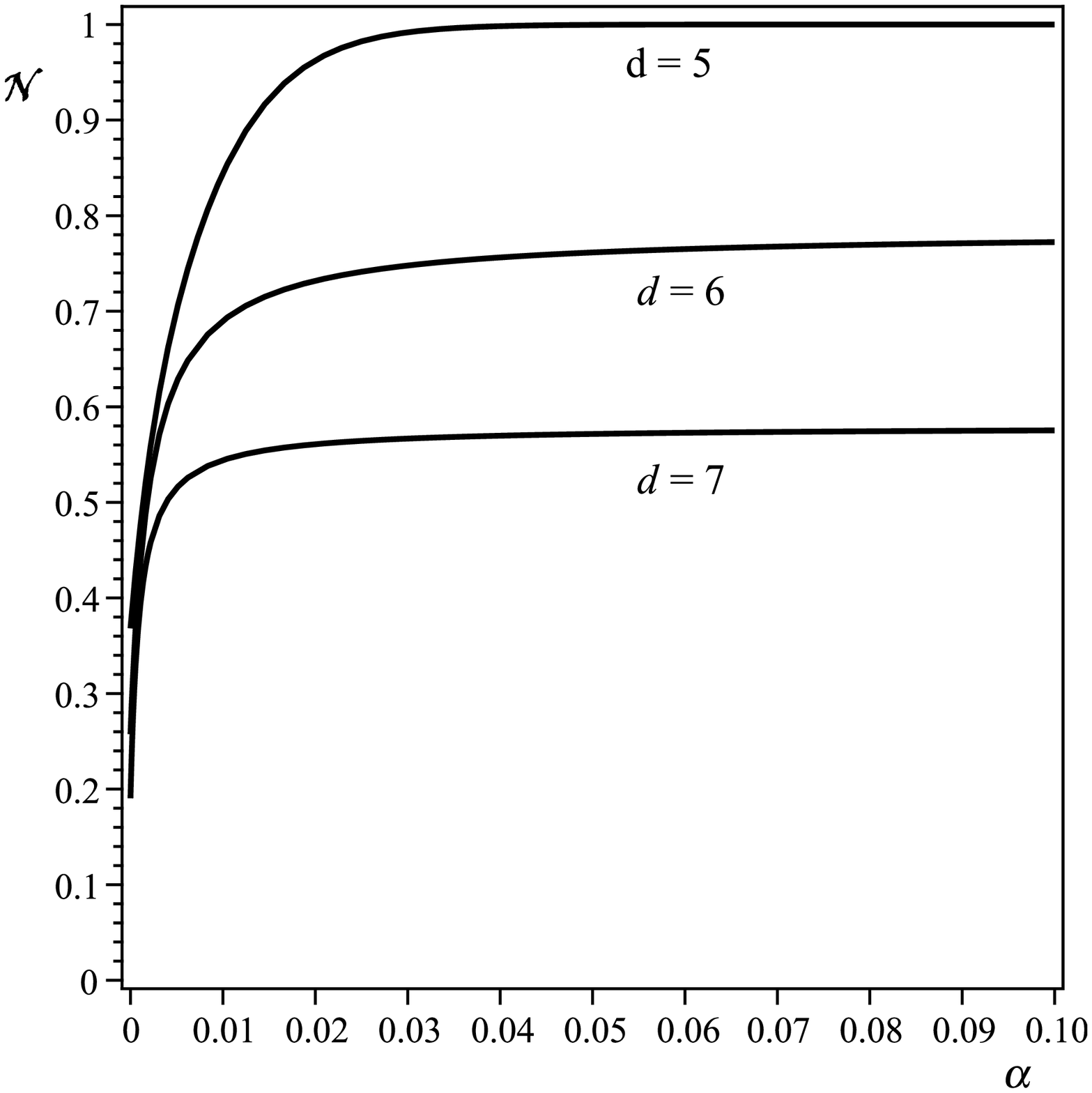}
    \end{minipage}}
\subfigure[ \,Logarithmic negativity versus the radius of the
horizon $r_h$ for some given $d$. There is a significant
difference between the curve for $d=5$ and the other curves. ]{
    \label{BLRR}     \begin{minipage}[b]{0.3\textwidth}
      \centering
      \includegraphics[width=5cm,height=5cm]{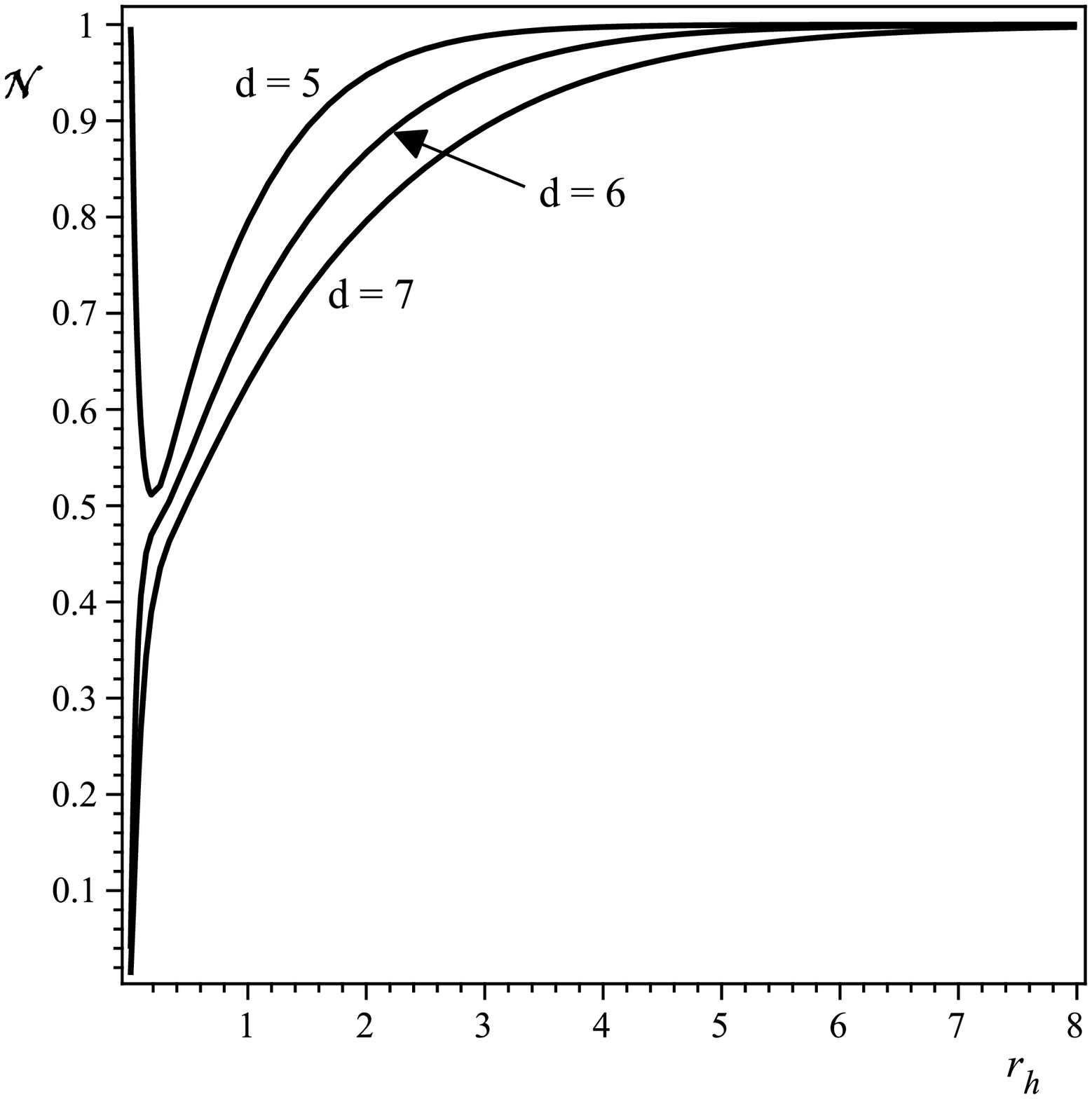}
    \end{minipage}}
\caption{ The logarithmic negativity for the bosonic entanglement.
} \label{BL}
\end{figure}

The curves in Fig. \ref{BL} describe the behavior of the
logarithmic negativity for the bosonic enatnglement. In Fig.
\ref{BLNT} the logarithmic negativity is plotted versus the
Hawking temperature $T$. We see that by increasing $T$, the
logarithmic negativity asymptotically descends to zero. The
behavior of the logarithmic negativity in terms of the spacetime
dimensions $d$ is shown in Fig. \ref{BLn} for some given $\alpha$.
We see that the logarithmic negativity decreases less by
increasing $\alpha$. This behavior is justified by recalling that
the positive Gauss-Bonnet coefficient leads to an antigravity
effect that can prevent the decoherence. In Fig. \ref{BLAlpha},
the $\alpha$-dependence of entanglement is shown. Only for $d=5$
the entanglement asymptotically reaches to the unity. In Fig.
\ref{BLRR}, the logarithmic negativity is plotted versus $r_h$,
the horizon radius. While in $d=5$, in a particular behavior, the
curve takes a minimum and return to unity for large radii, for
dimensions greater than five, the behavior is monotonic; the
entanglement grows from zero at $r_h=0$ and reaches asymptotically
to the unity. This difference can be justified by comparing Eqs.
(\ref{htd}) and  (\ref{ht5}). Especially note  that for $r_h=0$,
$T$ in (\ref{htd}) diverges , but $T$ in (\ref{ht5}) vanishes.

\subsection{Fermionic entanglement}

Suppose that the observers Alice and Rob take the Bell state
(\ref{bell}) which this time is built with the states of a Dirac
quantum field in the gravitational field. As the bosonic case we
assume that two observers are located first far from the black
hole horizon, where they share the Bell state. Then Alice falls
into the black hole freely, but Rob move toward the event horizon.
But he barely, decelerates and keep himself out of the horizon.
Using Eq. (\ref{Bogfer}), we can expand the Rob's kets in terms of
the Schwarzschild modes. Then,  the density operator of the system
can be obtained as
 \bea{ss}
       \rho_{\textrm{A,I,II}}&=&\frac{1}{2}\left(\cos^2\zeta\ket{000}\bra{000}
       +\sin^2\zeta\ket{011}\bra{011}+\ket{110}\bra{110}\right)\\
       \nonumber  &&+\frac{1}{2}\l \cos \zeta \sin \zeta
      \ket{000}\bra{011}+\cos \zeta \ket{000}\bra{110}+\sin
     \zeta\ket{011}\bra{110}+H. C. \r.
 \eea
This apparently describes a  tripartite system. However, Rob is
causally disconnected from the region II and so we must trace over
the states in that region, which results a mixed density matrix
denoted as $\rho_{\textrm{A},\textrm{I}}$. Then we obtain the
partially transposed of the resulting matrix as
\begin{equation}
     \rho^T_{\textrm{A},\textrm{I}}=\frac{1}{2}\left(
       \begin{array}{cccc}
          \cos ^{2}\zeta   &  0         &   0     &  0 \\
            0          & \sin ^2\zeta   & \cos \zeta  &  0 \\
            0          & \cos \zeta     & 0       &  0 \\
            0          &          0 &       0 &   1
       \end{array}
      \right).
\end{equation}
This has one negative eigenvalue $-\frac{\cos^2 \zeta}{2}$ which
obviously leads to  a logarithmic negativity as
 \be{LN1}
 \mathcal{N}=\log_2\l1+\cos^2 \zeta\r.
 \ee
Since $\tan{\zeta}=e^{-\pi\omega/T}$, $\zeta$ ranges from 0 to
$\frac{\pi}{4}$, then $\mathcal{N}$  can take only values between
$\thicksim 0.58$ and 1. This means that the fermionic entanglement
cannot be erased completely even for an infinite Hawking
temperature.

\begin{figure}
\subfigure[\,Logarithmic negativity versus the Hawking temperature.
The entanglement never reaches to the values less than $\thicksim
0.58$.]{
    \label{FLNT}   \begin{minipage}[b]{0.4\textwidth}
      \centering
      \includegraphics[width=5cm,height=5cm]{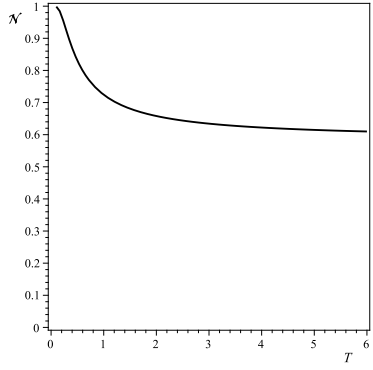}
    \end{minipage}}
 \subfigure[\, Logarithmic negativity versus the spacetime
dimensions for some given $\alpha$. By increasing $\alpha$, the
entanglement becomes less sensitive to increase of $d$.]{
    \label{FLd}  \begin{minipage}[b]{0.4\textwidth}
      \centering
      \includegraphics[width=5cm,height=5cm]{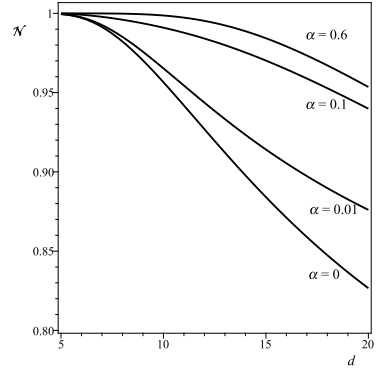}
    \end{minipage}}
\subfigure[\, Logarithmic negativity versus the Gauss-Bonnet
coefficient $\alpha$ for some given $d$. By increasing $\alpha$,
the entanglement grows; however, only for $d=5$ this can reach to
the maximal entanglement of 1.]{
    \label{FLAlpha}     \begin{minipage}[b]{0.4\textwidth}
      \centering
      \includegraphics[width=5cm,height=5cm]{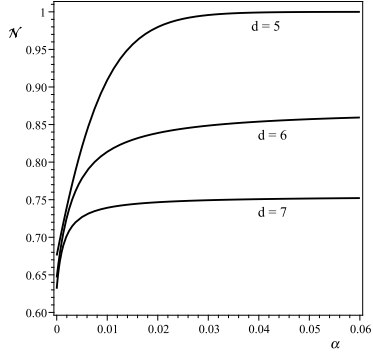}
    \end{minipage}}
\subfigure[\, Logarithmic negativity versus the radius of the
horizon $r_h$ for some given $d$. There is a significant
difference between the curve for $d=5$ and the other curves. ]{
    \label{FLR5D}     \begin{minipage}[b]{0.4\textwidth}
      \centering
      \includegraphics[width=5cm,height=5cm]{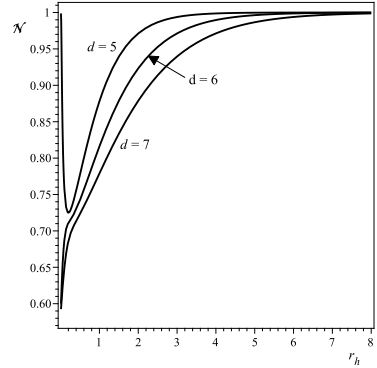}
    \end{minipage}}
\caption{ The logarithmic negativity for the fermionic
entanglement.}\label{FL}
 \end{figure}

In Fig. \ref{FL} we have plotted this logarithmic negativity
versus $T, d, \alpha$ and $r_h$. These curves are comparable with
the curves in Fig. \ref{BL}. The general behaviors of the
corresponding curves are the same; however, they differ in
details. Especially note to the difference between the asymptotic
or starting values of the curves in the corresponding figures. It
seems that the fermionic entanglement is generally robuster than
the bosonic entanglement.

\section{Fidelity of teleportation}\label{secFID}

\subsection{Teleportation by the bosonic field}

We set up the problem for performing a teleportation between the
observers Alice and Rob. Suppose that each cavity supports two
orthogonal modes with the same frequency labelled $A_i$ and $R_i$
with $i=1,\ 2$, such that each mode can be excited to a single
photon Fock state at the coincidence point. The state held by
Alice and Rob is chosen to be the entangled Bell state:
$$\ket{\Phi}=\frac{1}{\sqrt{2}}\l \ket{\mathbf{0}}_A\ket{\mathbf{0}}_R+\ket{\mathbf{1}}_A\ket{\mathbf{1}}_R \r,$$
where $\vert \mathbf{0} \rangle_{A}$, $\vert
\mathbf{1}\rangle_{A}$ are defined in terms of the physical Fock
states for Alice's cavity by the dual-rail basis states as
$\vert\mathbf{0} \rangle_{A}=\vert 1 \rangle_{A_{1}}\vert
0\rangle_{A_{2}}$, $\vert \mathbf{1}\rangle_{A}=\vert
0\rangle_{A_{1}}\vert 1 \rangle_{A_{2}}$, with  the similar
expressions for the Rob's cavity. Teleportation procedure provides
a  way to teleport an unknown state
$\ket{\mathbf{\Psi}}_A=\alpha\ket{\mathbf{0}}+\beta\ket{\mathbf{1}}$
to Rob, utilizing quantum entanglement. We should assume that
Alice has an additional cavity, which contains a single qubit with
dual-rail encoding by a photon excitation of a two-mode  vacuum
state. This will allow Alice to perform a joint measurement on the
two orthogonal modes of each cavity mandated with her. After
Alice's measurement, Rob becomes  aware of  this by a classical
channel. Rob's photon will be projected according to the
measurement outcome. The final state that Rob receives can be
given by $\vert \varphi_{ij} \rangle =x_{ij}\vert 0 \rangle
+y_{ij}\vert 1 \rangle$, where $(x_{00},y_{00})=(\alpha,\beta),
(x_{01},y_{01})=(\beta,\alpha),(x_{10},y_{10})=(\alpha,-\beta)$,
and $(x_{11},y_{11})=(-\beta,\alpha)$, are four possible
conditional state amplitudes. Once receiving the Alice's result of
measurement, Rob can apply a unitary transformation to verify the
protocol in his local frame. However, Rob must  notice the fact
that his cavity will become teemed with thermally excited photons
because of the Hawking-Unruh  effect and his state is  mixed. When
Alice sends the result of her measurement to Rob, if Alice has not
yet cross the  future horizon, the state that Rob observes must be
traced out over the region II, and the density state reduces to
\begin{eqnarray}
      && \rho^{(I)}_{ij}=\mathrm{Tr}_{II} (\vert \varphi_{ij}
       \rangle_{M} \langle \varphi_{ij}
      \vert)=\sum^{\infty}_{n=0}p_{n}\rho^{I}_{ij,n}
        \nonumber\\
        &&=\frac{1}{\cosh^{6}\eta}\sum^{\infty}_{n=0}\sum^{n}_{m=0}\left((\tanh^{2}\eta)^{n-1}[(n-m)|x_{ij}|^{2}+m
         |y_{ij}|^{2}]\times \vert m,n-m \rangle_{I} \langle m,n-m \vert
       \right.\\
        &&\left.+(x_{ij}y^{*}_{ij} \tanh^{2n}\eta \nonumber
     \sqrt{(m+1)(n-m+1)})\times \vert m,n-m+1\rangle_{I}
    \langle m+1,n-m \vert + {\rm H.C.}\right),
\end{eqnarray}
where $p_n$ denotes the  expansion coefficient given as
\begin{equation}
   \quad p_{0}=0, \quad p_{1}=1/\cosh^{6}\eta, \quad p_{n}=\frac{(\tanh^{2}\eta)^{n-1}}{\cosh^{6}\eta},
\end{equation}
and $\ket{m,n-m}$ denotes a state of $n$ total excitations in the
region I.

\begin{figure}
\subfigure[ \,Fidelity of teleportation versus the hawking
temperature. The curve rapidly drops to zero.]{
    \label{BTT}     \begin{minipage}[b]{0.4\textwidth}
      \centering
      \includegraphics[width=5cm,height=5cm]{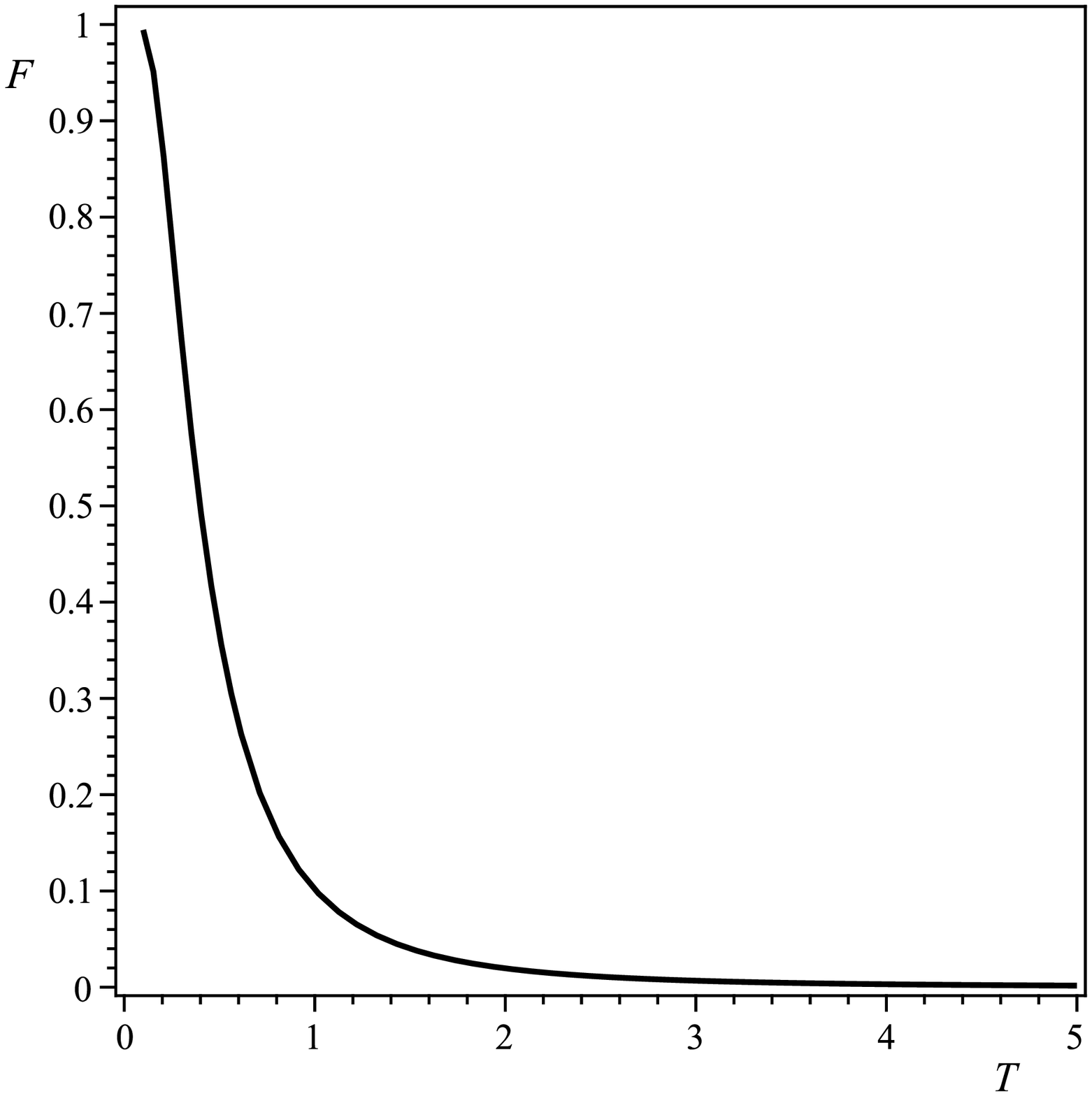}
    \end{minipage}}
\subfigure[\, Fidelity of teleportation versus the Gauss-Bonnet
coefficient $\alpha$ for some given $d$. Increase of $\alpha$
improves the teleportation.]{
    \label{TBAlpha}     \begin{minipage}[b]{0.4\textwidth}
      \centering
      \includegraphics[width=5cm,height=5cm]{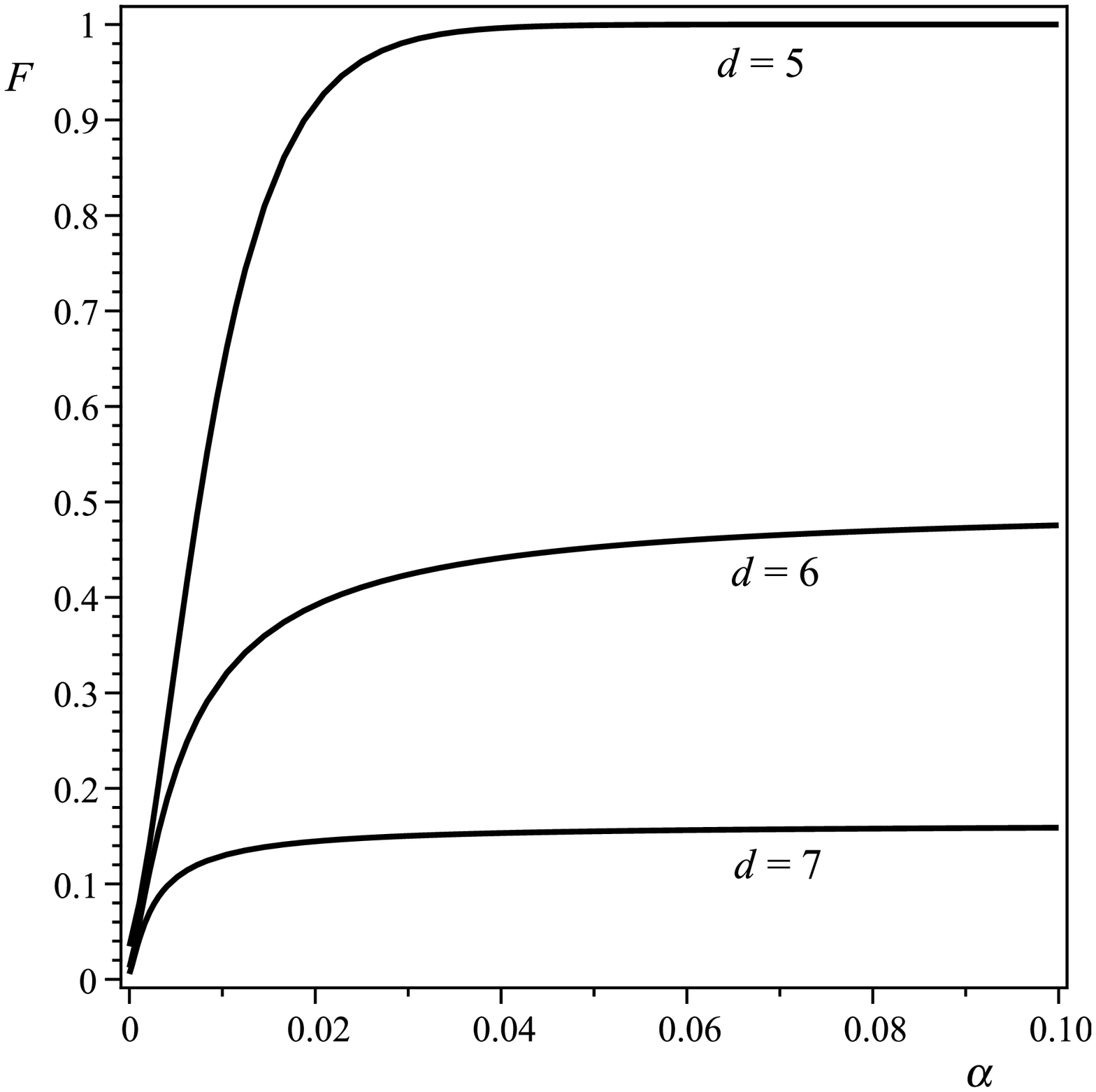}
    \end{minipage}}
\subfigure[ \,Fidelity of teleportation versus the spacetime
dimensions $d$ for some given $\alpha$. Additional dimensions lead
to more errors in the teleportation.]{
    \label{TBn}     \begin{minipage}[b]{0.4\textwidth}
      \centering
      \includegraphics[width=5cm,height=5cm]{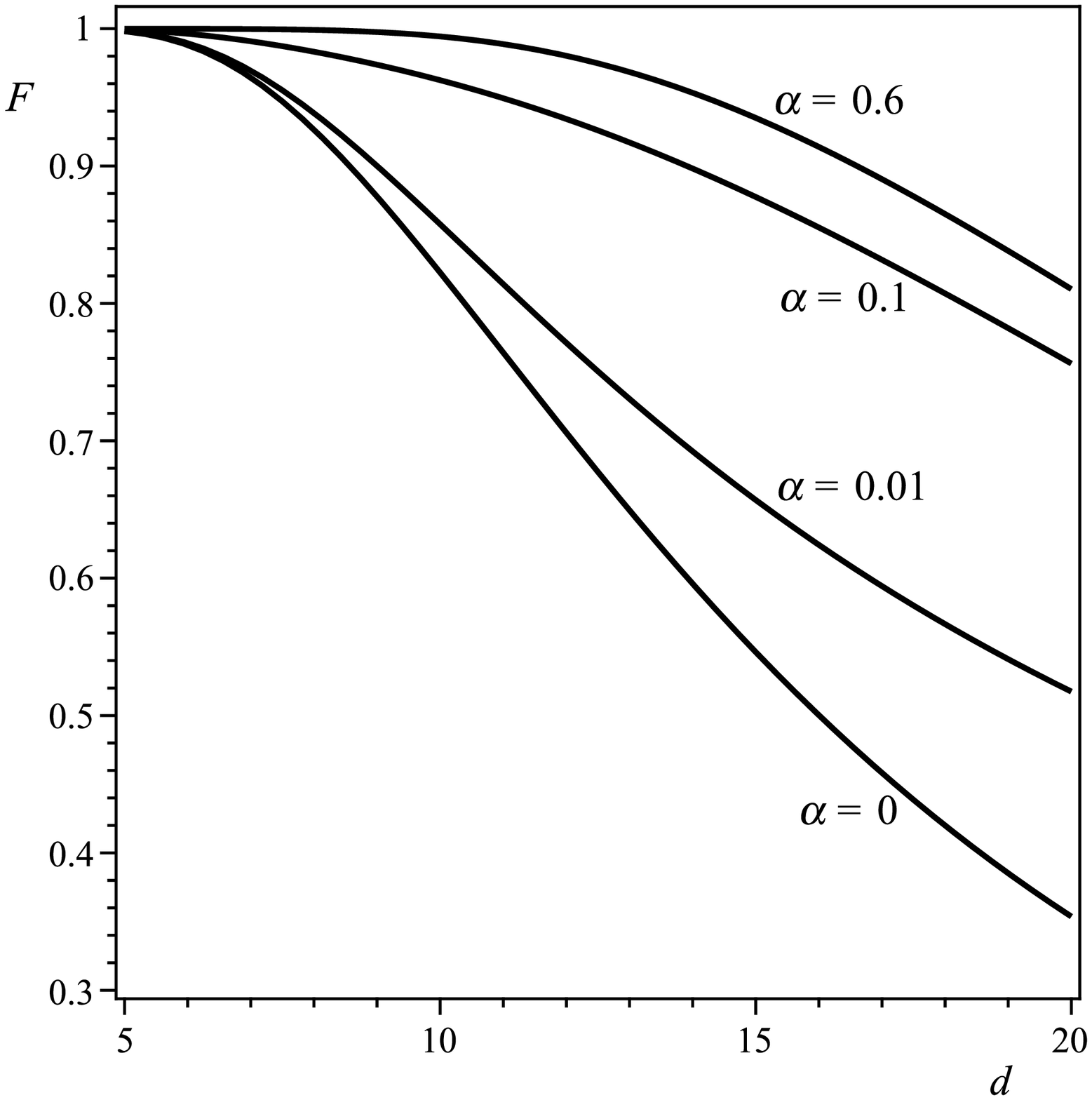}
    \end{minipage}}
\caption{ The fidelity of teleportation by the bosonic field.}
\label{TB}
 \end{figure}

Suppose that upon receiving the result $(i,j)$ of Alice's
measurement, Rob can apply the rotation operators restricted to
the 1-excitation sector of his state as $Z^i_IX^j_I|_{n=1}$. The
fidelity  of Rob's final state with the state that Alice attempts
to teleport to him is given by:
\begin{eqnarray}
    F\equiv
   \textrm{Tr}_{I}\l\ket{\mathbf{\Psi}}_I\bra{\mathbf{\Psi}}\rho^I\r=\cosh^{-6}\eta
\end{eqnarray}
In Fig. (\ref{TB}), we have plotted this fidelity in terms of $T$,
$\alpha$ and $d$. We see that by increasing $T$, the fidelity
rapidly falls to zero leading to a fatal error in the
teleportation. This means that the teleportation in this case is
very sensitive to the Hawking temperature. Also, it is seen that
the fidelity decreases in terms of spacetime dimension $d$, that
is, additional dimensions lead to more errors in the
teleportation. The fidelity increases by increasing $\alpha$, as a
consequence of antigravity role of $\alpha$.

\subsection{Teleportation  by the fermionic field}

\begin{figure}
 \subfigure[ \,Fidelity of teleportation versus the hawking
temperature. The curve never reaches to values less than 0.5.]{
    \label{FTT}     \begin{minipage}[b]{0.4\textwidth}
      \centering
      \includegraphics[width=5cm,height=5cm]{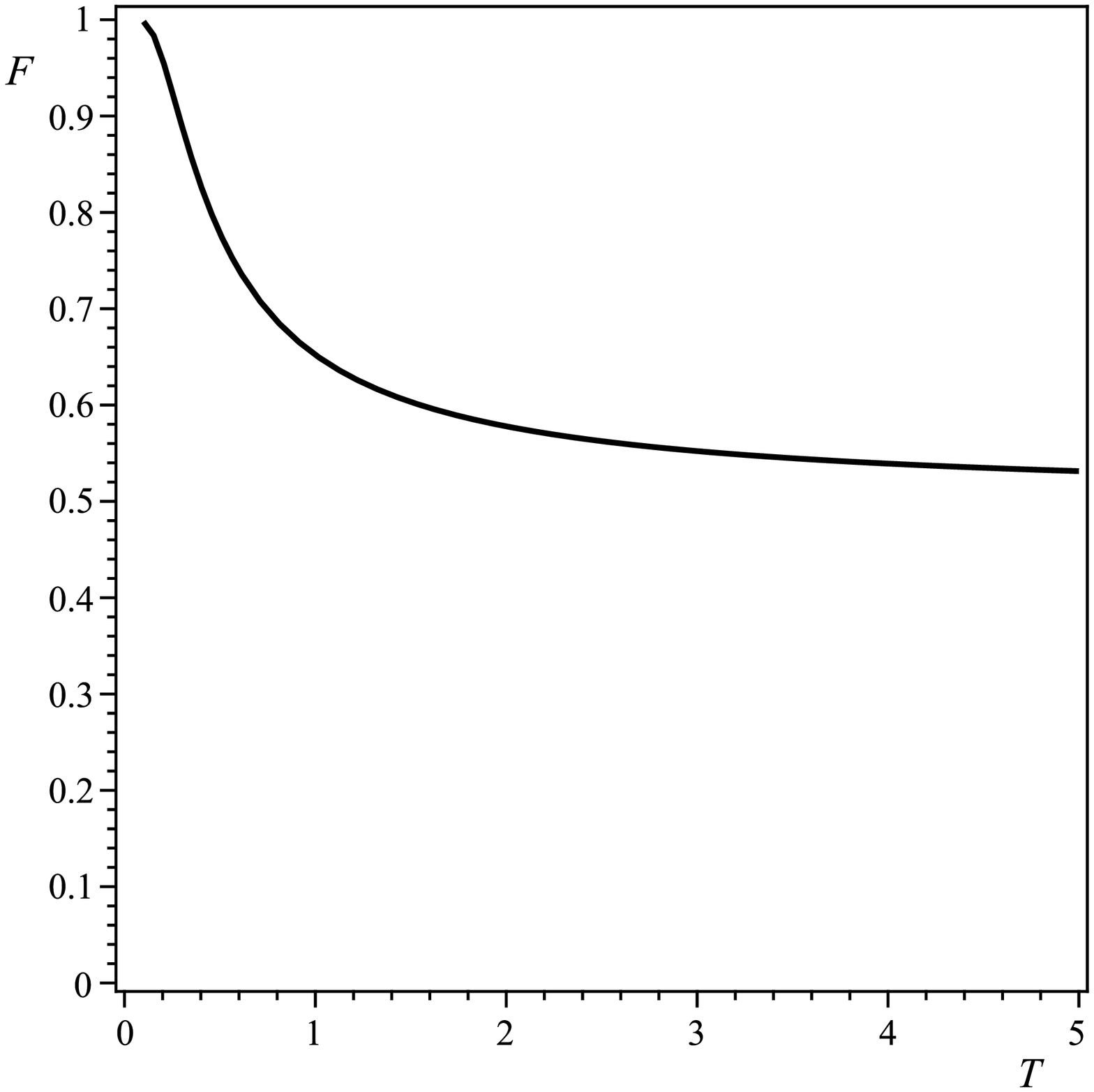}
    \end{minipage}}
\subfigure[\, Fidelity of teleportation versus the Gauss-Bonnet
coefficient $\alpha$ for some given $d$. Increase of $\alpha$
improves the teleportation. ]{
    \label{TFAlpha}     \begin{minipage}[b]{0.4\textwidth}
      \centering
      \includegraphics[width=5cm,height=5cm]{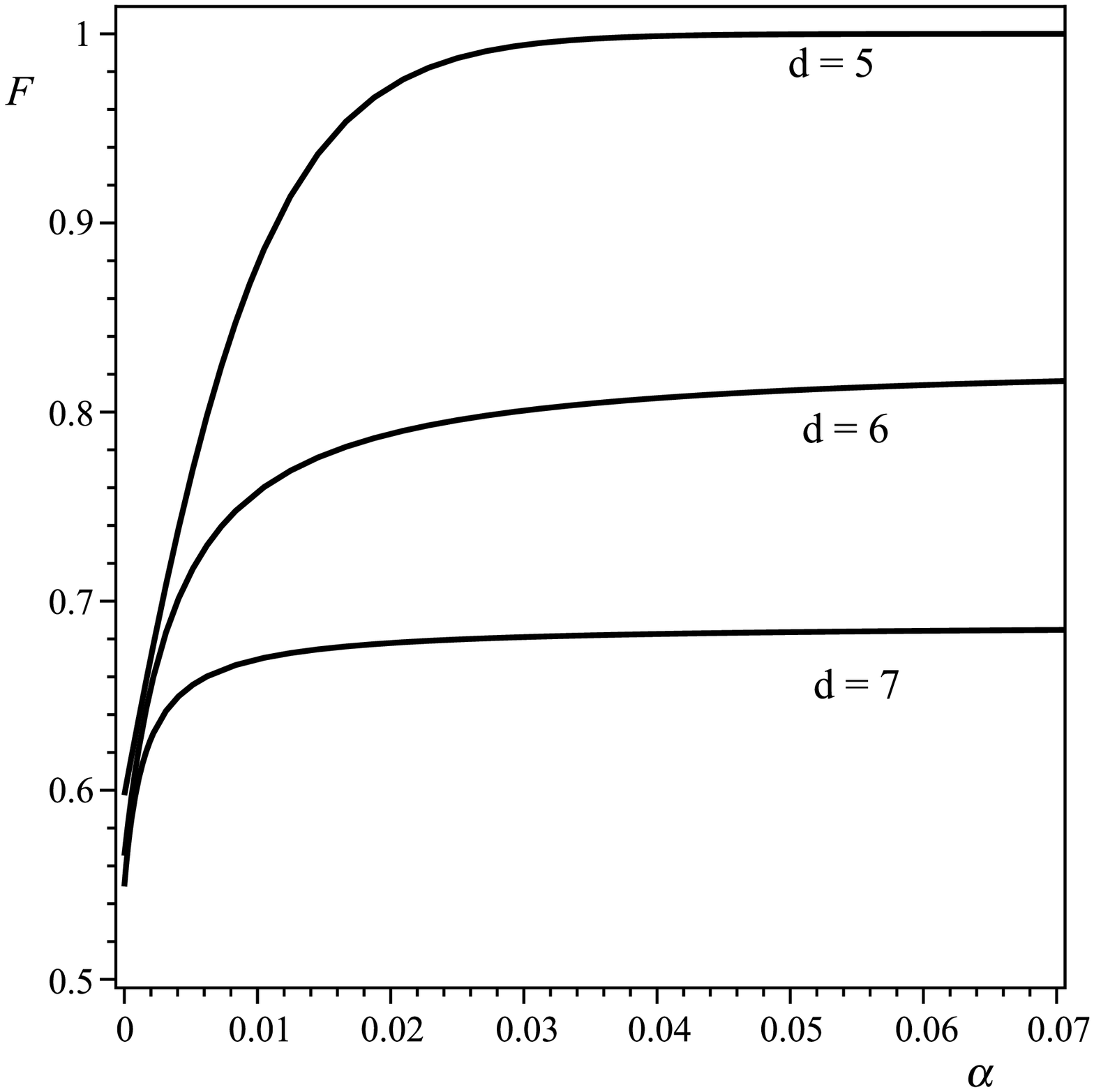}
    \end{minipage}}
\subfigure[\, Fidelity of teleportation versus the spacetime
dimensions $d$ for some given $\alpha$. Additional dimensions lead
to more errors in the teleportation. ]{
    \label{TFn}     \begin{minipage}[b]{0.4\textwidth}
      \centering
      \includegraphics[width=5cm,height=5cm]{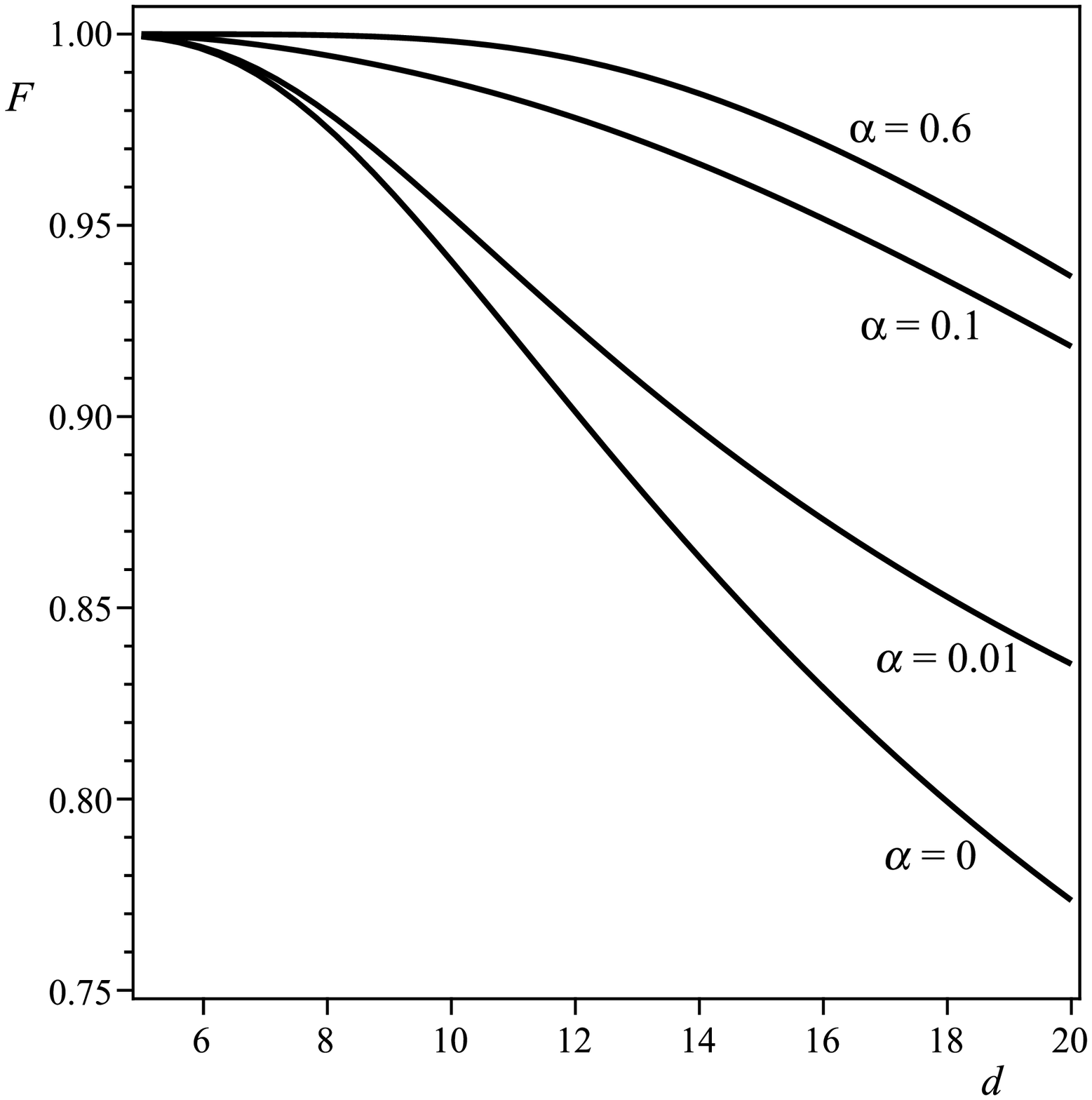}
    \end{minipage}}
    \caption{ The fidelity of teleportation by the fermionic field. } \label{TF}
\end{figure}

Using dual-rail basis as an excitation of a spin-up state in one
of two possible modes in Alice and Rob cavities, one obtains the
fidelity of Rob's final state as
      $$F=\cos^2\zeta$$
In Fig. (\ref{TF}) we have plotted this fidelity  versus $T,\,
\alpha$ and $d$. In this case the fidelity never reaches to values
less than 0.5. Although the general behavior of the fermionic
fidelity is similar to the bosonic fidelity, but the fidelity in
this case can not be destroyed completely, that is, in
teleportation by the fermionic field less errors occur. Again we
see that, in the Gauss-Bonnet gravitation theory the teleportation
can be improved.

\section{Conclusion}\label{fin}

In this work,  we studied the bipartite entanglement between the
states of a non-interacting bosonic or a fermionic field in the
spacetime of a $d$-dimensional spherically symmetric black hole of
the Einstein-Gauss-Bonnet gravitation. We considered two
observers; one of them, say Alice, was freely falling falling into
the black hole and so used an inertial frame described by the
Kruskal coordinates. The other observer, say Rob, accelerated to
avoid falling into the black hole, and so he was in a non-inertial
frame described by Schwarzschild-like coordinates. The Bogoliubov
transformations that relate the states in  the inertial frame to
the non-inertial frame, were calculated by comparing the problem
with its Schwarzschild analogue. We assumed that the observers
initially share a Bell state built by single the mode states of
the fields. We investigated the logarithmic negativity a measure
for the entanglement. Although general behavior of this bipartite
entanglement is similar to the behavior of the entanglement in the
Schwarzschild spacetime or entanglement in the accelerated Rindler
frames; however, some new features emerge here. In particular, the
Gauss-Bonnet coefficient $\alpha$, which determines the strength
of the higher derivative terms in the gravitation theory, and also
the suggested higher dimensions for the spacetime, lead to
important results. The Gauss-Bonnet term with positive $\alpha$
can play an antigravity role in the cosmological context. From the
viewpoint of quantum information theory this causes to decrease
the Hawking-Unruh effect and consequently reduces the entanglement
degradation. We showed that by increasing $\alpha$, the
logarithmic negativity saturated to values depending on the
dimension of the spacetime; only for $d=5$ this saturated to the
unity. Also, we studied the effect of higher dimensions on the
entanglement degradation. By increasing the dimensions of the
spacetime more entanglement degradation occurs. Moreover, we
investigated the behavior of the logarithmic negativity in terms
of the radius of the horizon. For $d=5$, the logarithmic
negativity starts from 1 at $r_h=0$, takes a minimum at a given
$r_h$, and finally return to 1. However, in other dimensions this
grows uniformly from zero (for the bosonic field), or a nonzero
value (for the fermionic field), to the unity.

As an application of these results, we discussed the teleportation
between Alice and Rob, who use modes of the bosonic or the
fermionic field. We calculated the fidelity as a quantitative
measure of the accuracy of the information transmission.
Expectedly, the behavior of the fidelity is in complete agreement
with the behavior of the logarithmic negativity.

A possible extension of this work includes the black hole
solutions of higher order Lovelock gravitation theory. Then we
encounter more parameters that affect on the quantum information
process.

\end{spacing}
\end{document}